# THE STRONG COUPLING CONSTANT FROM THE LATTICE 3-GLUON VERTEX

D. HENTY, C. PARRINELLO

*Dept. of Physics & Astronomy, University of Edinburgh, Edinburgh EH9 3JZ, U. K.*

C. PITTORI*

*L.P.T.H.E., Université de Paris Sud, Centre d'Orsay, 91405 Orsay, France*

We compute the QCD running coupling on the lattice as defined from the 3-gluon vertex. We present the results of an exploratory study at $\beta = 6.0$ on a $16^4$ lattice, which show that for momenta larger than 2 GeV, the coupling runs according to the 2-loop asymptotic formula, allowing a precise determination of the $\Lambda$ parameter. Our renormalization procedure corresponds to a momentum subtraction scheme in the continuum and, most remarkably, one does not need lattice perturbation theory to match the results to $\overline{MS}$ scheme. We obtain $\alpha_s^{\overline{MS}}(M_Z) = 0.115 \pm 0.003 \pm 0.008$.

*Talk presented by C.Pittori

## 1 Introduction

We report here the preliminary results of a recent lattice determination of the QCD running coupling, $\alpha_s$, from the 3-gluon vertex [1]. This method differs from the previous ones as the coupling is determined in terms of the fundamental fields and vertex functions of the continuum Lagrangian. This is achieved by evaluating two and three-point off-shell gluon Green functions, in the Landau gauge, and imposing non-perturbative renormalization conditions directly on them [2]. By varying the renormalisation scale, the method allows the determination of $\alpha_s(q)$ for different momenta from a single simulation. Of course this is valid provided that the renormalisation scales can be chosen in a range of values between the non-perturbative region and the high momentum region, where discretization effects become important. An analogous non-perturbative approach has been recently applied to the renormalisation of composite fermion operators [3]. As will be explained in the following, a crucial feature of this procedure, which corresponds to a momentum subtraction renormalisation in continuum QCD, is that renormalised Green's functions do not depend on the lattice regularisation. As a consequence, lattice perturbation theory (LPTH) is not needed in order to relate the measured coupling to $\alpha_s^{\overline{MS}}$, and the relation between the two schemes can be entirely computed in continuum perturbation theory. Despite recent proposals [4] to improve the convergence of LPTH series, this step still provides one important source of systematic errors in the final prediction for $\alpha_s^{\overline{MS}}$, expecially at low $\beta$ values. Finally, given the conceptual simplicity of the method, the extension to unquenched QCD presents in principle no extra problems.

## 2 The Method

Provided systematic errors are under control, in order to compute $\alpha_s^{\overline{MS}}$ from the lattice it is sufficient to:
1) fix the scale of momenta by determining the lattice spacing $a$;
2) choose a suitable renormalisation scheme which defines the renormalised coupling;
3) match the result to the $\overline{MS}$ scheme.

As for step 1) we take the value of $a^{-1}$ determined by Bali and Schilling [5] in their string tension measurements, that is $a^{-1} = 1.9 \pm 0.1$ GeV at $\beta = 6.0$. We quote a systematic error on the scale to take into account the uncertainty resulting from other possible choices.

In our renormalisation scheme, step 2), the running coupling at scale $\mu$ is defined from the renormalised amputated 3-gluon vertex, as

$$g_R(\mu) \equiv \frac{\Gamma_R^{(3)}(p/\mu, 0, -p/\mu)}{2p_\beta}\Big|_{p^2=\mu^2}. \quad (1)$$

On the lattice the amputated bare 3-gluon vertex is defined by evaluating first the complete 3-point function $G_B^{(3)}{}_{\alpha\beta\gamma}(ap_1, ap_2, ap_3)$ and then *amputating* its external legs by dividing it by the gluon self-energies corresponding to the external momenta. We work in the Landau gauge and choose the asymmetric kinematical point defined by

$$\alpha = \gamma, \qquad p_1 = -p_3 = p_\beta, \ p_2 = 0. \quad (2)$$

This is a choice suitable for the lattice symmetries, and such that the tensor structure of the vertex reproduces the tree-level one [2]. Then in the $a \to 0$ limit, the bare 3-gluon vertex can be written as

$$\Gamma^{(3)}_{B\,\alpha\beta\alpha}(pa,0,-pa) = 2\, Z_V^{-1}(pa)\, g_0(a)\, p_\beta. \qquad (3)$$

where $g_0$ is the bare coupling constant. Notice that on the lattice the calculation for $\Gamma^{(3)}_B$ yields directly the product $Z_V^{-1} g_0$ in a non-perturbative way. The renormalised 3-gluon vertex is than related to the bare one through (the $a \to 0$ limit is understood in the following)

$$\Gamma^{(3)}_R(p/\mu)|_{p^2=\mu^2} = Z_A^{3/2}(\mu a)\Gamma^{(3)}_B(pa)|_{p^2=\mu^2}, \qquad (4)$$

where $Z_A$ is the gluon wave function renormalisation. In order to compute $Z_A$, we impose that for a fixed momentum scale $p^2 = \mu^2$ the renormalised propagator takes its continuum tree-level value. Thus the renormalisation condition is simply

$$G^{(2)}_R(p/\mu)|_{p^2=\mu^2} = Z_A^{-1}(\mu a)G^{(2)}_B(pa)|_{p^2=\mu^2}$$
$$= G^{TREE}(\mu) = \frac{1}{\mu^2}, \qquad (5)$$

where $G^{(2)}_B(pa)$ is the scalar part of the bare gluon self-energy in the Landau gauge. The above equation defines $Z_A$ in a non-perturbative way.

In this way one can express $g_R(\mu)$ entirely in terms of two and three-point bare Green's functions non-perturbatively evaluated on the lattice

$$g_R(\mu) = \frac{|p|^3}{2p_\beta} \frac{G^{(3)}_{B\,\alpha\beta\alpha}(pa,0,-pa)}{G^{(2)}_B(0)\sqrt{G^{(2)}_B(ap)}}|_{p^2=\mu^2}. \qquad (6)$$

The above formula is the basis of our computational procedure.

## 3 Numerical Results

The lattice calculation was performed on a 16K Connection Machine 200. We generated two data sets, each comprising 150 gauge configurations, on a $16^4$ lattice, at $\beta = 6.0$ and 6.2. The $\beta = 6.2$ lattices were produced essentially for testing purposes, so that in this paper we only quote numerical results obtained at $\beta = 6.0$ A crucial step in the method is the accurate implementation of the lattice Landau gauge condition by using a Fourier-accelerated algorithm.

The calculation of the renormalised running coupling then proceeds as follows. As an initial step, we evaluate the gluon propagator and compute $Z_A$ from eq.(5). Next, we compute the complete three-point function $G^{(3)}_B$ of the gluon field. Finally, $g(\mu)$ is obtained from eq.(6). This is plotted as a function of $\mu = |p|$ in Fig.(1).

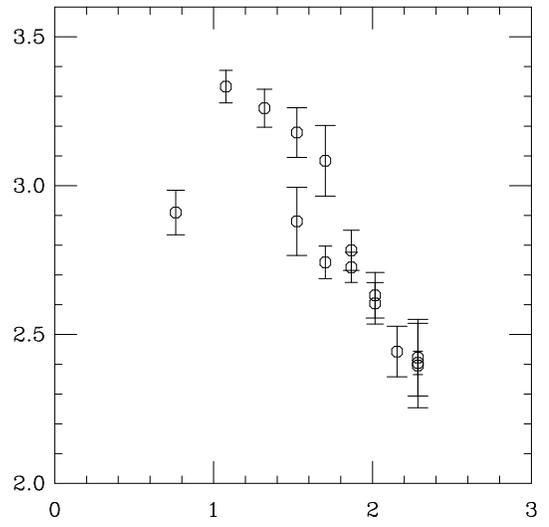

Figure 1: Running coupling vs. momentum.

The tree-level form of a bosonic lattice propagator suggests that higher order terms in the lattice spacing are small when the lattice momenta satisfy $p = 2\pi\, n/La \ll \pi/2a$, where $n$ is an integer. As our lattice size is $L = 16$, we limit lattice momenta values to $n \leq 3$ in each direction. In addition, we will also discard momenta such that $p^2 > (6\pi/La)^2$. The important physical question is whether there exists a range of momenta for which the coupling that we measure runs according to the 2-loop asymptotic formula. To answer this question, we compute $\Lambda_{\widetilde{MOM}}$ as a function of the measured values of $g(\mu)$ according to

$$\Lambda_{\widetilde{MOM}} = \mu\, exp\left(-\frac{1}{2b_0 g^2(\mu)}\right)\left[b_0 g^2(\mu)\right]^{-\frac{b_1}{2b_0^2}}, \qquad (7)$$

where $b_0 = 11/16\pi^2$, $b_1 = 102/(16\pi^2)^2$ and $\Lambda_{\widetilde{MOM}}$ is the QCD scale parameter for the renormalization scheme that we are using. In the asymptotic region, we expect $\Lambda_{\widetilde{MOM}}$ defined from the above equation to be a constant.

We plot $\Lambda_{\widetilde{MOM}}$ versus $q$ in Fig.(2); it appears that for $q > 2$ GeV the data are consistent with a constant value for the QCD scale parameter. In order to extract from the above data a prediction for $\Lambda_{\widetilde{MOM}}$, we fit the data points for $\mu > 2\,GeV$ to a constant. This yields

$$\Lambda_{\widetilde{MOM}} = 0.96 \pm 0.02 \pm 0.09\; GeV, \qquad (8)$$

where the first error is statistical and the second one comes from the uncertainty on the value of $a^{-1}$.

## 4 Matching to $\overline{MS}$

In order to extract from the numerical results for $\Lambda_{\widetilde{MOM}}$ a prediction in the $\overline{MS}$ scheme - step 3 in section 2 - we only need the ratio $\Lambda_{\overline{MS}}/\Lambda_{\widetilde{MOM}}$, which can be computed entirely in continuum perturbation theory.

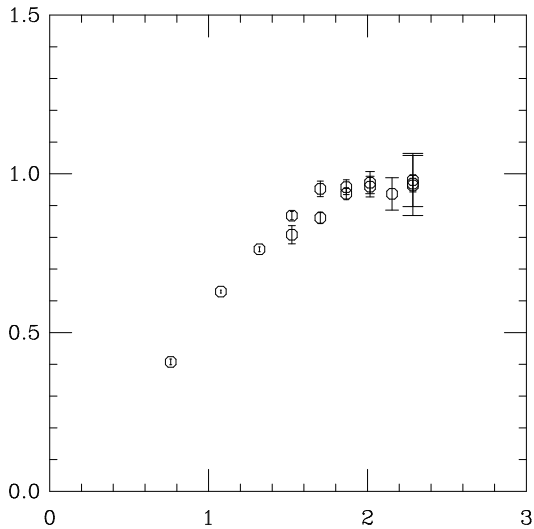

Figure 2: Λ parameter vs. momentum.

The general relation between the Λ's in two schemes $A$ and $B$, can be written as

$$\frac{\Lambda_A}{\Lambda_B} = \exp\left[-\frac{1}{2\beta_0}\left(\frac{1}{g_A^2(\mu)} - \frac{1}{g_B^2(\mu)}\right) + O(g^2(\mu))\right]. \quad (9)$$

In our case, in the Landau gauge and in the quenched approximation, $n_f = 0$, we get

$$\frac{\Lambda_{\overline{MS}}^{n_f=0}}{\Lambda_{\widetilde{MOM}}^{n_f=0}} = 0.35 \quad (10)$$

Note that the scheme that we have adopted in the non-perturbative calculation differs from the usual $\widetilde{MOM}$ scheme as it contains an extra constant term in the vertex renormalization constant [1]. This is a linear term in the momenta, not proportional to the tree level, equal on the lattice and in the continuum, which is perturbatively included in (10). Using the nonperturbative result (8), we get

$$\Lambda_{\overline{MS}}^{n_f=0} = 340 \pm 10 \pm 40 \ MeV \quad (11)$$

This is the main result of our computation. We emphasise that even for $\beta$ as low as 6.0, the uncertainty from the one-loop matching condition is smaller than the error on the determination of the scale and comparable to the size of the statistical error. Of course this depends on the fact that we only use continuum perturbation theory, and it would not be true with other methods which rely on LPTH.

At this point we face the problem of extracting a prediction for the full theory. In particular, we aim to estimate $\alpha_s^{\overline{MS}}(M_Z)$. One can take into account, at least partially, the effect of quenching, by assuming that the quenched and the 3-flavor couplings would be approximately the same at scales as low as $\sim 0.7 \ GeV$. A different procedure to estimate the effect of quenching is described in ref.[1]. By averaging the two results and evolving the coupling up to the $M_Z$ scale, we would obtain

$$\alpha_s^{\overline{MS}}(M_Z) = 0.115 \pm 0.003 \pm 0.008 \quad (12)$$

where the first error is dominated by the uncertainty on $a^{-1}$, and the second one keeps into account the residual effect of the quenched approximation. This result is compatible with the first unquenched calculations, see [6], and with other quenched lattice computations [7].

## 5 Conclusion

We have shown that a non-perturbative determination of the QCD running coupling can be obtained from first principles by a lattice study of the triple gluon vertex. Systematic lattice effects appear to be under control in our calculation. In ref.[1] a comparison with a corresponding LPTH calculation provides some important consistency checks. In the future we plan to study further the role of infrared and ultraviolet cutoffs, by increasing the physical volume and the range of $\beta$ values for our lattices.

## Acknowledgments

D. Henty and C. Parrinello acknowledge the support of PPARC through a Personal Fellowship (DH) and grant GR/J 21347 (CP). C. Pittori acknowledges the support of an HCM Fellowship ER-BCHBICT930887.